\documentclass[aps,prl,twocolumn,floatfix,superscriptaddress,showpacs,amsfonts
,amssymb,amsmath,preprintnumbers]{revtex4-1}
\usepackage{bm}
\usepackage{epsfig}
\usepackage{graphicx}
\usepackage{color}
\usepackage{subfigure}

\newcommand{\fig}[1]{Fig.~\ref{#1}}
\newcommand{\Fig}[1]{Fig.~\ref{#1}}
 
\newcommand{\eq}[1]{Eq.~(\ref{#1})}

\newcommand{\Epar}{E_X}
\newcommand{\Eparmax}{E_X^{\rm max}}

\newcommand{\be}{\begin{equation}}
\newcommand{\ee}{\end{equation}}

\newcommand\bea{\begin{eqnarray}}
\newcommand\eea{\end{eqnarray}}

\renewcommand{\d}{\partial}
\newcommand{\Apar}{A_{\parallel}}

\newcommand{\Apareq}{A_{\parallel eq}}
\newcommand{\Aparzero}{A_{\parallel 0}}
\newcommand{\vpar}{v_{\parallel}}
\newcommand{\vthe}{v_{{\rm th}e}}
\newcommand{\Tpar}{T_{\parallel e}}

\newcommand{\kperp}{k_\perp}

\newcommand\lapperp{\nabla^2_\perp}
\newcommand{\rhoi}{\rho_{i}}
\newcommand{\rhos}{\rho_s}

\newcommand{\nuei}{\nu_{ei}}
\newcommand{\Dprime}{\Delta'}

\newcommand{\hypercoll}{\nu_{coll}}
\renewcommand{\(}{\left(}
\renewcommand{\)}{\right)}
\renewcommand{\[}{\left[}
\renewcommand{\]}{\right]}

\begin{document}

\title{Fast collisionless reconnection and electron heating in strongly magnetized plasmas}
\author{N.\ F.\ Loureiro}
\affiliation{Associa\c{c}\~ao EURATOM/IST, Instituto de Plasmas e Fus\~ao Nuclear --- 
Laborat\'orio Associado,\\
Instituto Superior T\'ecnico, Universidade T\'ecnica de Lisboa, 1049-001 Lisboa, Portugal}
\author{A.\ A.\ Schekochihin}
\affiliation{Rudolf Peierls Centre for Theoretical Physics, University of Oxford, 
Oxford OX1 3NP, UK}
\author{A.\ Zocco}
\affiliation{EURATOM/CCFE Fusion Association, Culham Science Centre, Abingdon OX14 3DB, 
UK}
\affiliation{Rudolf Peierls Centre for Theoretical Physics, University of Oxford, 
Oxford OX1 3NP, UK}

\date{\today}

\begin{abstract}
Magnetic reconnection in strongly magnetized (low-beta), weakly collisional plasmas is 
investigated
using a novel  fluid-kinetic model [Zocco \& Schekochihin, Phys. Plasmas {\bf 18}, 102309 (2011)]
which retains non-isothermal electron kinetics.
It is shown that electron heating via Landau damping (linear phase mixing) 
is the dominant dissipation mechanism. 
In time, electron heating occurs after the peak of the reconnection rate; in space, it
is concentrated along the separatrices of the magnetic island.
For sufficiently large systems, the peak reconnection rate is 
 $cE_{max}\approx 0.2v_AB_{y,0}$, 
where $v_A$ is the Alfv\'en speed based on the reconnecting field $B_{y,0}$.
The island saturation width is the same as in MHD models except for small systems, 
when it becomes comparable to the kinetic scales. 

\end{abstract}


\maketitle


\paragraph{Introduction.}
Magnetic reconnection is a 
reconfiguration of the magnetic field in a plasma via
localized unfreezing of the magnetic flux~\cite{Zweibel_09}.
It is commonly associated with energy release 
in astrophysical and laboratory plasmas, 
but the details of the energy conversion mechanisms and partition between 
particle species and fields are poorly understood.
This Letter focuses on a key 
aspect of this issue: 
the conversion of the magnetic energy into 
electron internal energy, i.e., electron heating.

We consider a fundamental reconnection paradigm: 
the tearing mode in a periodic box~\cite{FKR}. The tearing instability leads
to the opening, growth and 
saturation of a magnetic island~\cite{FKR,Rutherford_73,Escande_04,Militello_04,Loureiro_05}.
Since the saturated width of the island is in general macroscopic, it cannot depend on the 
microphysics of the plasma and, in particular, should not depend on its collisionality 
(as we will show).
This implies that the total fraction of the initial magnetic energy converted 
to other forms of energy 
from the begining to the end of the evolution of the tearing mode must be the 
same in collisional and collisionless plasmas.
In a periodic (closed) system, energy cannot be lost
via bulk plasma outflows. Therefore, the magnetic energy difference between the initial 
and final states
must be accounted for by conversion into the thermal energy of the particles. 
In collisional plasmas, this is achieved by
Ohmic and viscous heating~\cite{Loureiro_12}. 
However, many natural systems where reconnection occurs are only 
weakly collisional;
the only available heating channels then are Landau damping and electron viscosity, 
both of which ultimately rely on the electron collision frequency being finite, though possibly 
arbitrarily small.
In this Letter, we show that electron heating via linear phase mixing associated with Landau damping
is the main energy conversion channel in weakly collisional reconnection in strongly 
magnetized (low-beta) plasmas.

\paragraph{Equations.}
We use a fluid-kinetic approximation applicable to low-beta plasmas 
(``KREHM''~\cite{ZS_11}).
Its main distinctive feature, for our purposes, is the coupling of 
Ohm's law to the electron (drift) kinetic equation via the electron temperature fluctuations.
The kinetic equation allows for both collisions and Landau resonance, thus enabling ``collisionless''
electron heating.

We work in the low-$\beta$ regime, where $\beta$, the ratio of the plasma to the magnetic pressure,
is ordered similar to the electron-ion mass ratio $m_e/m_i$. 
Let us define the perturbed electron distribution function to lowest 
order in $\sqrt{m_e/m_i}\sim\sqrt\beta$,
and in the gyrokinetic expansion~\cite{Frieman_Chen_82,Krommes_02,Howes_06,Abel_12} as 
$\delta f_e= g_e + (\delta n_e/n_{0e}+2\vpar u_{\parallel e}/\vthe^2)F_{0e}$,
where $F_{0e}$ is the equilibrium Maxwellian,   
$\vthe = \sqrt{2 T_{0e}/m_e}$ is the electron thermal speed (with $T_{0e}$ the mean 
electron temperature), $\vpar$ is the parallel 
velocity coordinate, 
$\delta n_e/n_{0e}$ is the electron density perturbation 
(the zeroth moment of $\delta f_e$) normalized 
to its 
background value $n_{0e}$,
and $u_{\parallel e}=(e/c m_e)d_e^2\nabla_\perp^2\Apar$ is the parallel electron 
flow (the first moment of
$\delta f_e$; $\Apar$ is the parallel component of the vector potential 
and $d_e=c/\omega_{pe}$ is the 
electron skin depth). 
All moments of $\delta f_e$ higher than $\delta n_e$ and $u_{\parallel e}$
are contained in 
$g_e$, e.g., the electron temperature perturbation is
$\delta \Tpar/T_{0e}=(1/n_{0e})\int d^3{\bf v}~(2\vpar^2/\vthe^2)~g_e$.
The dynamics of the plasma is described by the evolution equations for 
$\delta n_e/n_{0e},~\Apar$ and $g_e$, which, for 
the 2D case considered here, read~\cite{ZS_11}:
\begin{align}
\label{eq:cont}
&\frac{1}{n_0}\frac {d \delta n_e}{dt}= 
\frac{1}{B_z}\left\{\Apar,\frac{e}{c m_e}d_e^2\nabla_\perp^2 \Apar\right\},\\
\label{eq:Ohm}
&\frac {d}{dt} \(\Apar - d_e^2\nabla_\perp^2 \Apar\)= 
-\frac{c T_{0e}}{eB_z}
\left\{\Apar,\frac{\delta n_e}{n_{0e}}+\frac{\delta \Tpar}{T_{0e}}\right\},\\
\label{ge_eq}
&\frac{dg_e}{dt} -
\frac{\vpar}{B_z}\left\{\Apar, g_e-\frac{\delta \Tpar}{T_{0e}}F_{0e}\right\} = C[g_e]\nonumber\\
&\qquad\qquad-\(1-\frac{2\vpar^2}{\vthe^2}\)
\frac{F_{0e}}{B_z}\left\{\Apar,\frac{e}{cm_e}d_e^2\nabla_\perp^2\Apar\right\},
\end{align} 
where $C[g_e]$ is the collision operator~\footnote{We keep the collision operator in 
\eq{ge_eq}, but ignore the resistive term $\eta\nabla_\perp^2\Apar$ on the RHS of \eq{eq:Ohm}.
This is formally inconsistent, since 
$\eta = \nuei d_e^2$; in practice, the only effect of setting $\eta=0$ is allowing the use of 
larger values of $\nuei$ than would otherwise be required in order to access the 
collisionless regime.}, 
$\{...,...\}$ is the Poisson
bracket and
$d/dt=\d/\d t + c/B_z\{\varphi,...\}$,
with $B_z$ the out-of-plane magnetic guide-field and $\varphi$ the electrostatic potential.
The latter is obtained via the gyrokinetic Poisson's law~\cite{Krommes_02}, 
$\delta n_e/n_{0e}=1/\tau (\hat\Gamma_0-1) e \varphi/T_{0e}$,
where $\tau=T_{0i}/T_{0e}$ and $\hat\Gamma_0$ is the real-space operator that is 
the inverse Fourier transform of 
$\Gamma_0(\alpha)=I_0(\alpha)e^{-\alpha}$, with $I_0$ the
modified Bessel function and $\alpha=\kperp^2\rho_i^2/2$ ($\rhoi=v_{{\rm th} i}/\Omega_i$ is the 
ion Larmor radius).

\eq{ge_eq} shows that the popular isothermal closure~\cite{Schep_94}, 
$g_e=0$, is not 
a solution of that equation
unless $\{\Apar, \nabla_\perp^2\Apar\}=0$, a condition that cannot describe a reconnecting  
system [though it does describe the (macroscopic) island 
saturation~\cite{Escande_04, Militello_04}, as we will find].
This means that at least the possibility of electron heating in weakly collisional reconnection
cannot be ignored.
\paragraph{Numerical Details.}
To simplify the solution of \eq{ge_eq}, note 
that it does not contain an explicit dependence  
on the perpendicular velocity coordinate, $v_\perp$.
If we ignore any such a dependence that is introduced by the collision operator, 
$v_\perp$ can be integrated out, so $g_e=g_e(x,y,\vpar, t)$.
Next, we introduce the Hermite  expansion
$g_e(x,y,t,\vpar)=\sum_{m=2}^\infty H_m(\vpar/\vthe) g_m (x,y,t)F_{0e}(\vpar)/\sqrt{2^m m!}$ 
($g_0=g_1=0$ because $\delta n_e$ and $u_{\parallel e}$ have been explicitly separated in 
the decomposition of $\delta f_e$ adopted above).
\eq{ge_eq} then unfolds into a series of coupled, fluid-like 
equations for each of the coefficients $g_m$: 
\begin{align}
\label{gm_eq}
&\frac{d g_m}{d t} = \frac{\vthe}{B_z}\(\sqrt{\frac{m+1}{2}}\left\{\Apar, g_{m+1}\right\}+
\sqrt{\frac{m}{2}}\left\{\Apar,g_{m-1}\right\}\)\nonumber\\
&\quad
+\frac{\sqrt{2}}{B_z}\delta_{m,2}\left\{\Apar,\frac{e}{c m_e}d_e^2\lapperp\Apar\right\}
-\hypercoll m^4 g_m,
\end{align}
where we have adopted a model (hyper) collision operator with 
$\hypercoll = 1/(\Delta t M^{4})$, where $M$ is the index of 
the highest Hermite polynomial kept in a simulation, and $\Delta t$ the 
timestep. 
Thus, in our simulations, 
$M$ is a proxy for the collision frequency, 
with higher values of $M$ corresponding to less collisional systems
(at the large values of $M$ reported here, $\Delta t\sim M^{-1/2}$ so $\hypercoll\sim M^{-7/2}$).
Note that the more familiar (and more physical) in such 1D problems 
L\'enard-Bernstein collision operator would instead be $m\nu_{ei}g_m$~\cite{ZS_11}
($\nu_{ei}$ is the electron-ion collision frequency);
because of the linear dependence on $m$ of this operator an unfeasibly large 
number of Hermite polynomials would need to be kept in order to resolve the velocity-space cutoff 
as $\nu_{ei}\rightarrow 0$. This is why we use hyper-collisions instead.

Eqs. (\ref{eq:cont},\ref{eq:Ohm},\ref{gm_eq}) are solved numerically using 
a pseudo-spectral code~\cite{Loureiro_08}.
The spatial grid size is $384^2$.
The resolution in velocity space is set by $M$ and ranges from $30$ to $500$.
Hyper-diffusive terms of the form 
$\nu_H \nabla_\perp^6$, where 
$\nu_H=0.25/\Delta t (\Delta x/\pi)^6$, 
with $\Delta x$ the grid spacing, 
are added to the RHS of all equations (including~\eq{gm_eq}).
These are  required to prevent the unbounded thinning of the current layer in the nonlinear 
regime~\cite{Ottaviani_93,Rogers_96,Valori_00}, but will not, as we will discover, 
dissipate much energy.
Physically, they stand in for collisional and collisionless electron finite Larmor radius 
effects (which are formally small in the KREHM approximation). 

The hyper-diffusive and hyper-collisional terms are the only dissipative terms we employ.
We shall distinguish their effects:
``hyper-viscous'' and ``hyper-Ohmic'' refer, respectively, to the dissipation terms 
in \eq{eq:cont} and 
in \eq{eq:Ohm}; ``Landau'' refers to the dissipation arising via 
both hyper-collisions and hyper-dissipation in \eq{gm_eq}. 
The rationale for the latter is that the only way that energy can arrive at any 
$g_m$ with $m>2$ is via phase mixing. Once there, how exactly it dissipates is not important: 
what we wish to investigate is the relative importance of dissipation via phase mixing 
(i.e., Landau-damping) {\it vs.} 
(hyper-) viscosity or resistivity.

The equilibrium in-plane magnetic field is  $B_{y,eq}=-d\Apareq/dx$, with
$\Apareq=\Aparzero/\cosh^2(x/a)$, where $a$ is the (normalizing) 
equilibrium scale length.
The (normalizing) Alfv\'en time is defined as $\tau_A=a/v_A$ with $v_A$ 
the Alfv\'en speed based on $B_{y,0}=\max|B_{y,eq}|=1$.
The simulations are performed in a doubly periodic box of dimensions $L_x\times L_y$, with 
$L_x/a=2\pi$ and $L_y$ such that $\hat k_y= 2\pi a/L_y$ yields the desired value of 
the tearing instability parameter 
$\Dprime a =2(5-\hat k_y^2)(3+\hat k_y^2)/(\hat k_y^2\sqrt{4+\hat k_y^2})$~\cite{Loureiro_05}.
For simplicity and numerical convenience, we set $\rhoi=\rhos=d_e=0.25a$, where 
$\rhos=\rhoi/\sqrt{2\tau}$. Thus, this study does not address any effects associated 
with the scale separation between ions and electrons --- a reasonable first step for 
low-$\beta$ plasmas.

\paragraph{Reconnection Rate.}
\begin{figure}
  \includegraphics[width=\columnwidth]{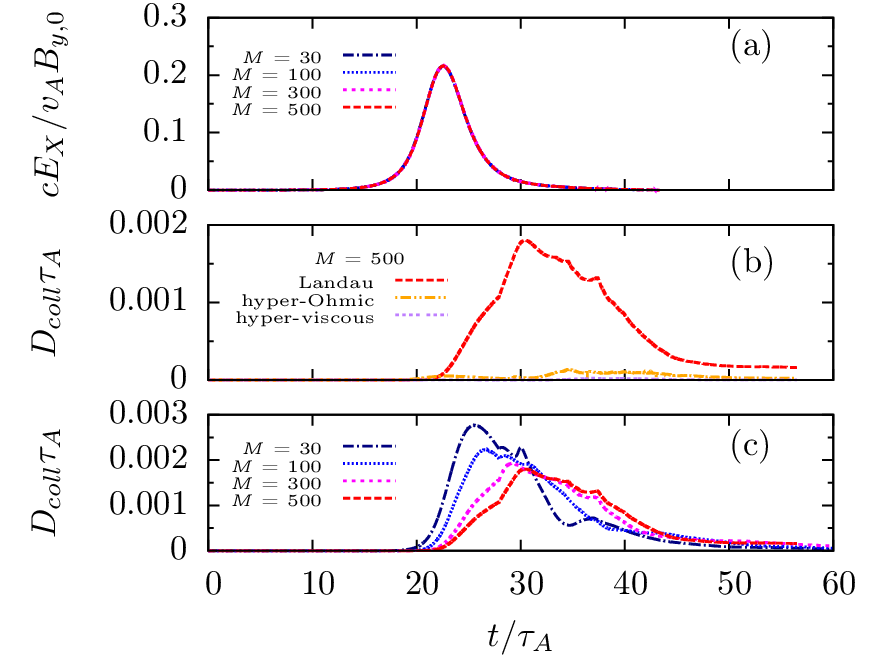}
\caption{Time traces of: (a)  the reconnection rate for different values of collisionality 
(represented by $M$; larger $M$ means less collisions); (b) 
the rates of dissipation via Landau damping and hyper-dissipation, for the least 
collisional case ($M=500$); (c) the rate of dissipation via
Landau damping for different values of $M$.
All runs had $\Dprime a=20$.}
\label{fig:rec_rate}
\end{figure}
Before we discuss the electron heating, let us first report some of the more standard features 
of our simulations. 
The system evolves in time from the linear to the nonlinear stage; saturation occurs when all the 
initially available flux has been reconnected.
The time-traces of the reconnection rate,
defined as the value of the parallel electric field at $(x,y)=(0,0)$ (the $X$-point), 
$\Epar$, are plotted in \fig{fig:rec_rate}(a).
As shown, the reconnection rate is entirely independent of collisionality (parameterized by $M$),
consistent with the fact that our simulations are in the weakly-collisional regime, 
where the frozen-flux constraint is broken by electron inertia, not the collisions.
The maximum value is $c\Eparmax\approx0.22 v_AB_{y,0}$, similar to the fast reconnection rates
obtained in the opposite limit of weak guide-field~\cite{Birn_01}, and in qualitative
agreement with~\cite{Rogers_01,rogers_07}.
Regarding the dependence of $\Eparmax$ with system size (not shown), we found that $\Eparmax$  
increases with $\Dprime$, asymptoting to $c\Eparmax\sim 0.2 v_A B_{y,0}$ for $\Dprime a\gtrsim 10$.

\begin{figure}
  \includegraphics[width=9cm]{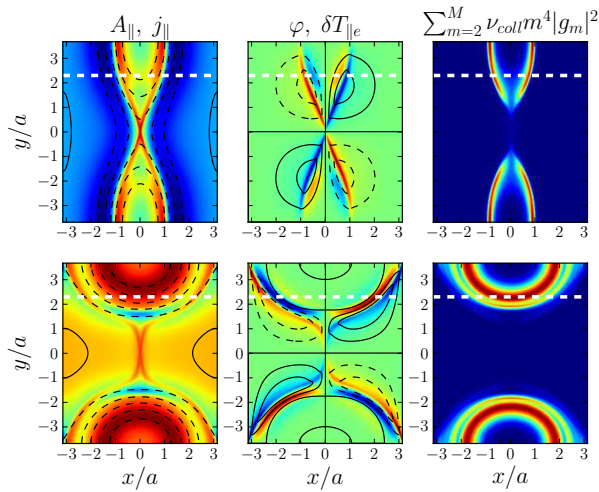}
  \vskip-0.25cm
\caption{System configuration for our least collisional simulation ($M=500$) at the time of 
maximum reconnection rate (top row)
and of maximum dissipation rate (bottom row) 
[see Figs.~\ref{fig:rec_rate}(a) and (b), respectively]. 
From left to right, the plots show: $\Apar$ (lines: full/dashed are positive/negative contours) 
and $j_{\parallel}=-c/4\pi\lapperp\Apar$ [colors, ranging from blue (negative) to red (positive)]; 
$\varphi$ (lines) and $\delta T_{\parallel e}$ (colors); the total dissipation via $g_m$'s
(i.e., via Landau damping).
The horizontal dashed line marks the location of the domain cut 
where the distribution function 
of \fig{fig:spectrum} (bottom row) is plotted.}
\label{contour}
\end{figure}
\Fig{contour} depicts the system configuration for our least collisional simulation ($M=500$) at
the time of maximum reconnection rate [$t/\tau_A=22.7$; see~\fig{fig:rec_rate}(a)] (top row) 
and at the time of maximum dissipation rate [$t/\tau_A=29.4$; see~\fig{fig:rec_rate}(b)] 
(bottom row).
A typical $X$-point geometry is seen, 
accompanied by a quadrupole structure exhibited by $\delta T_{\parallel e}$ 
(and by $\delta n_e$, not shown)~\cite{Uzdensky_06}~\footnote{Note that in none of 
our runs did we observe the 
current sheet becoming unstable to plasmoid formation, though this may be because even our 
largest system is in fact not large enough: the longest box is $L_y/a=3.785\pi$ 
(for $\Dprime a=54$); 
this run yields the maximum ratio of the current sheet length to $\rho_s$ of $\sim 8$, 
too small to expect a transition to multiple X-line reconnection~\cite{Ji_Daughton_11, Huang_11}.}.

\begin{figure}
   \includegraphics[width=8cm]{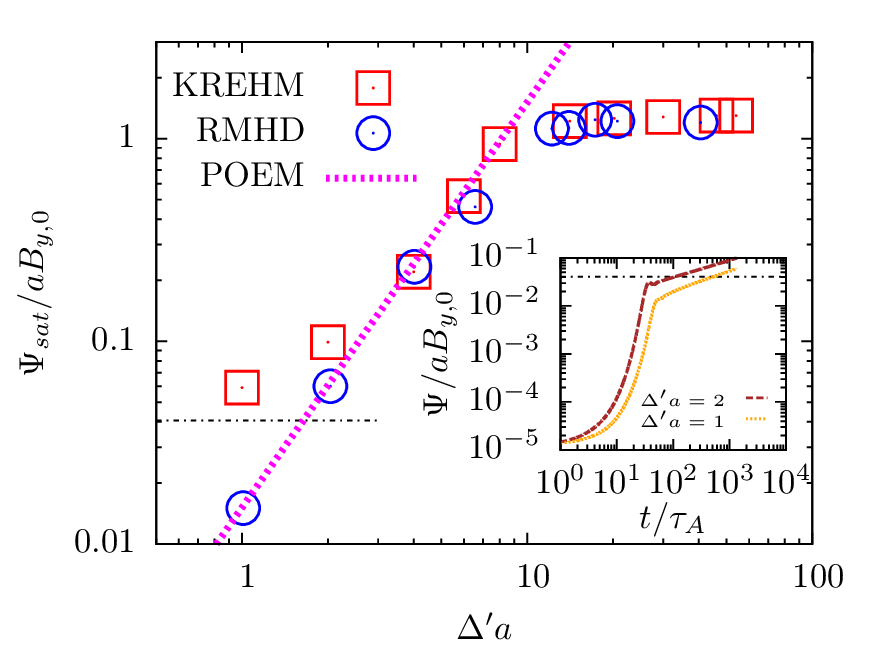}
\caption{Saturated flux $\Psi_{sat}$ {\it vs.} $\Delta'$.
``KREHM'' are the data from the kinetic simulations;
``RMHD'' are MHD results from~\cite{Loureiro_05}.
The dotted line (``POEM'') is the prediction from MHD 
theory~\cite{Escande_04,Militello_04}.
Inset: time-traces of flux for $\Delta'a=(1,2)$.
In both plots, the horizontal line is the saturated flux corresponding 
to a full island width of $2d_e/a$.}
\label{fig:wsat}
\end{figure}

\paragraph{Saturation.}
\fig{fig:wsat} shows the saturation amplitudes obtained in our simulations
(``KREHM''); overplotted are the MHD ($\rho_{s,i}=d_e=0$) results from 
Ref.~\cite{Loureiro_05} (``RMHD''). 
Also shown is the prediction 
from MHD theory~\cite{Escande_04,Militello_04} (``POEM''), valid in the small-$\Dprime a$ regime.
As anticipated, we find that saturation in our weakly collisional, kinetic simulations is 
well described by the (isothermal) MHD model.
At small $\Dprime a$, this agreement breaks down because the island saturation amplitude 
becomes comparable to the kinetic scales ($d_e,~\rho_{s,i}$). 
Saturation then becomes a slow, diffusive 
process, as illustrated by the inset plot: the islands slowly expand until their 
width becomes $\sim d_e$.
Thus, there is a lower limit 
to the saturation amplitude, set by $d_e$ (cf.~\cite{Sydora_01}); 
indeed, the frozen-flux condition precludes the 
definition of magnetic field lines at sub-$d_e$ scales.

\paragraph{Electron heating.}
We have established that the amount of energy converted during the evolution of the system is 
independent of the collisionality; we have also checked (not shown) that 
for all but the smallest systems, the energy 
converted into electron heating is a significant fraction of the initial (magnetic) 
energy, reaching $\sim 60\%$ for the largest systems.
We now turn to the main focus of this Letter:
how is the energy converted.
\fig{fig:rec_rate}(b) shows  the time traces of the dissipation rates 
for $\Dprime a =20$ and $M=500$.
We see that dissipation happens almost exclusively via the phase mixing/Landau 
channel. We reiterate that, {\it a priori}, the system is free to choose between different 
dissipation channels, i.e., the fact that phase mixing is the preferential dissipation mechanism 
is not hard-wired and is, therefore, a remarkable demonstration of the dominance of 
Landau damping over other dissipation mechanisms in weakly collisional
reconnection~\footnote{If we use a (formally incorrect)
isothermal closure, most of the dissipation occurs via Ohmic heating.}.

Another noteworthy feature in \fig{fig:rec_rate}(b) is the time lag between the 
peaks of the reconnection and dissipation rates.
This implies that magnetic energy is not directly dissipated 
by the reconnection process itself~\footnote{This is why the reconnection rate should be 
independent of the details of the dissipation: indeed, the same rate should be obtained even in  
dissipation-free Hamiltonian models~\cite{Kuvshinov_94,Grasso_01,Tassi_08,Waelbroeck_09,Sarto_11}, 
although they would not obtain the 
same energy partition that we report here.}.
Instead, it is first converted to other forms and then dissipated.
Indeed Landau damping dissipates the electron free energy, 
$\int dxdy/V\int d\vpar T_{0e}g_e^2/(2F_{0e})$.
\fig{fig:rec_rate}(c) shows that the time lag increases weakly with decreasing collisionality
(see below).

\begin{figure}
\vskip6.5cm
  \includegraphics[width=9cm]{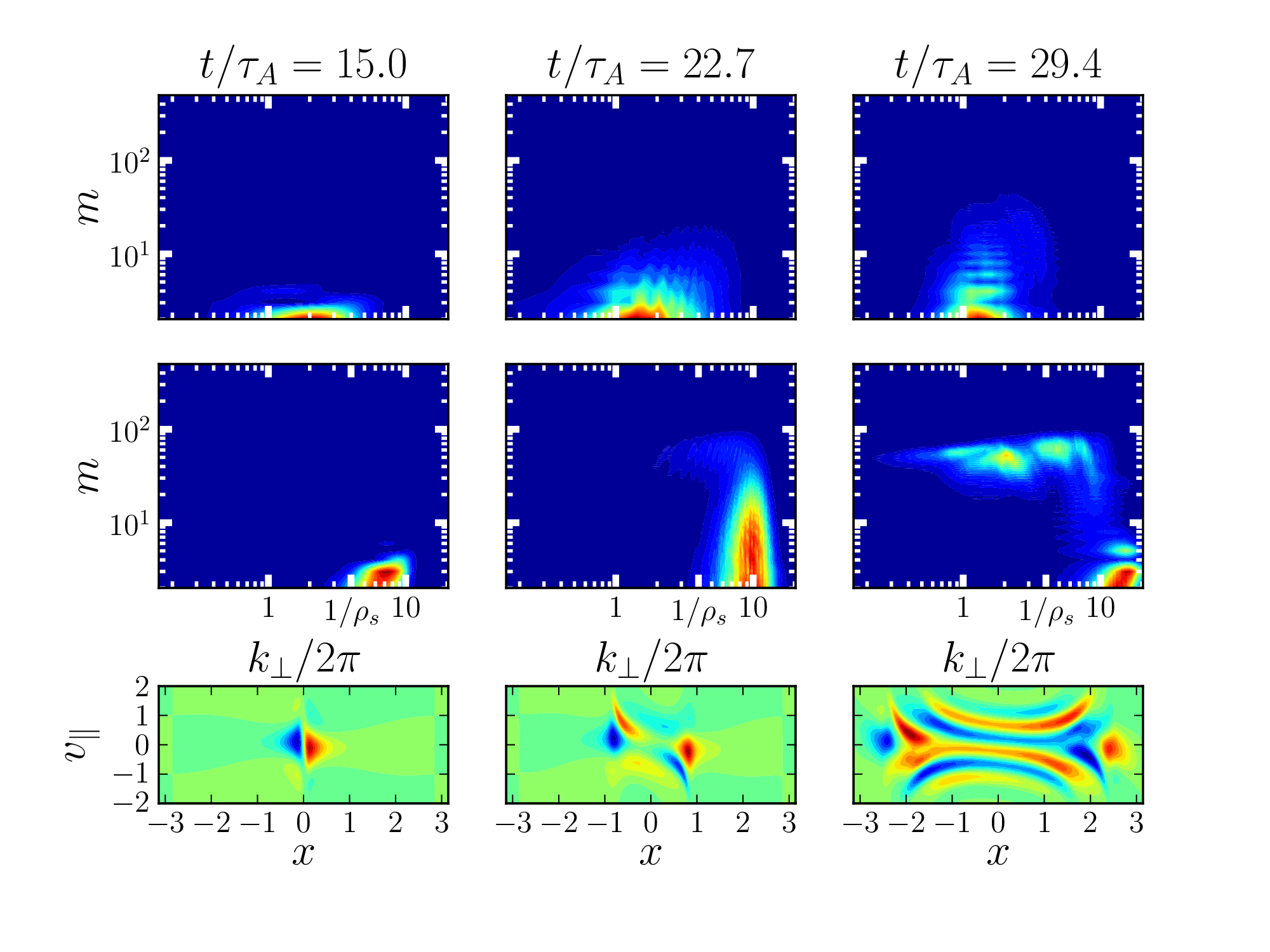}
\vskip-0.5cm
  \caption{Electron free energy spectrum (top row), dissipation 
spectrum (middle row) and a cut at $y/a=2.3$ (cf.~\fig{contour}) of the 
distribution function (bottom row) at the early nonlinear stage 
(left) and at the peaks of the reconnection rate (center) and dissipation rate (right)
for a run with $\Dprime a=20$ and $M=500$.}
  \label{fig:spectrum}

\end{figure}

A related question is where in the domain the heating is occurring. 
The plots in the right column 
of \fig{contour} show that there is no significant heating in the current sheet; 
instead, it happens predominantly along the separatrices of the island. 
This is consistent with the 
existence of a time lag between reconnection and dissipation:
the magnetic energy that is converted in the current sheet is mostly channeled into ion and 
electron kinetic energy. Both species flow downstream predominantly along the separatrices.
Deceleration of these flows converts kinetic energy into the free energy (or entropy) of the 
electrons (plotted in the top row of \fig{fig:spectrum}). 
It is this conversion that constitutes the energetics of Landau damping and results eventually 
in electron heating.

A detailed understanding of the electron heating process is yielded by the 
spectral maps [in the two-dimensional Fourier-Hermite ($\kperp,m$) space]
of the electron free energy and its dissipation. These are shown in the two
top pannels of \fig{fig:spectrum} for $M=500$. 
As time advances, the electron free energy cascades
to higher values of $m$, corresponding to the formation of small scales in velocity space, 
i.e., phase mixing. 
During this process, energy dissipation occurs via the hyper-diffusive terms, acting at large values 
of $\kperp$ (see middle pannel, second row). 
At later times, large enough values of $m$ are reached so the collisional 
dissipation becomes important and, indeed, dominant; 
this $m$-cutoff is clearly visible in the rightmost plot of the 
second row of \fig{fig:spectrum}. 

\begin{figure}
  \includegraphics[width=8cm]{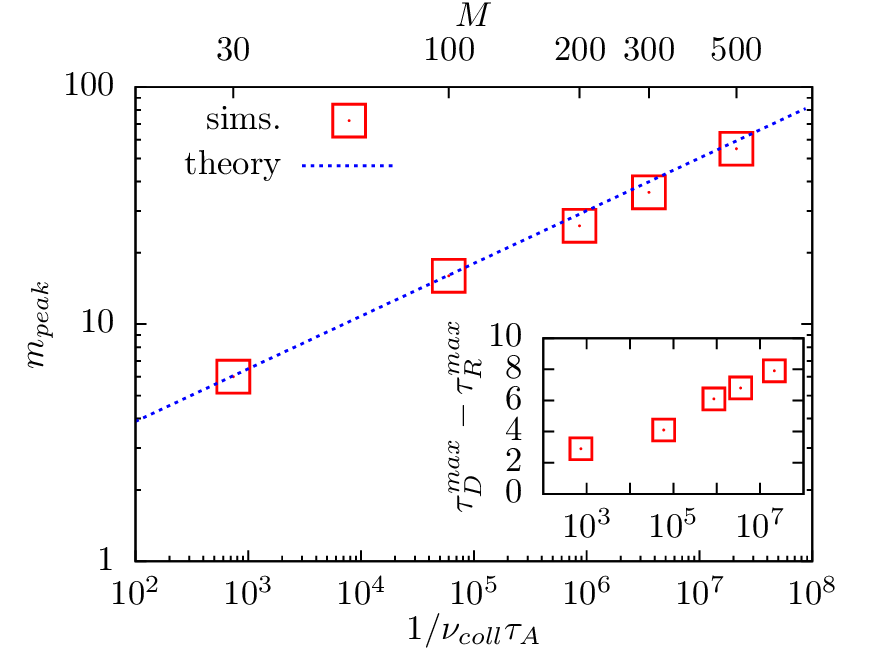}
  \caption{Value of $m$ at which most energy is dissipated, $m_{peak}$, as a function of 
collisionality, $\hypercoll$.
Inset: time lag between the peak of the reconnection rate ($\tau_R^{max}$) and the peak of the 
dissipation rate, ($\tau_D^{max}$), as a function of $\hypercoll$.}
\label{fig:mpeak_scaling}
\end{figure}

Let us estimate the velocity-space dissipation scale.
Linearising \eq{gm_eq} for a given $k_y$ we find that the 
electron free energy spectrum $E_m=|g_m|^2/2$ satisfies~\cite{ZS_11}
\be
\frac{\d E_m}{\d t} = -|k_y|\frac{B_y}{B_z} \vthe\frac{\d}{\d m}\sqrt{2m}E_m
-2\hypercoll m^4E_m.
\ee
Setting $\d E_m/\d t=2\gamma E_m$, we obtain
\be
\label{eq:E_m}
E_m = \frac{C(k_y)}{\sqrt{m}}\exp\[-\(\frac{m}{m_\gamma}\)^{1/2}-\(\frac{m}{m_c}\)^{9/2}\],
\ee
where $m_\gamma=[k_y B_y/B_z\vthe/(2\sqrt{2}\gamma)]^2$,
$m_c=[9/(2\sqrt{2}) k_y B_y/B_z \vthe/\hypercoll]^{2/9}$ 
and $C(k_y)$ is some $k_y$-dependent constant. 
As is evidenced by \fig{fig:rec_rate}(a), $\gamma$ is independent of collisions and 
thus so is $m_{\gamma}$.
Therefore, while the mode is strongly growing, $m_\gamma<m_c$ and so 
the collisional cuttoff cannot be reached.
This explains why the peak of the dissipation rate must occur later than that 
of the reconnection rate.
As reconnection proceeds into the saturation regime,
$\gamma\rightarrow 0$, so, regardless of how small $\hypercoll$ is, eventually 
$m_\gamma > m_c$ and, from then onwards, the Hermite spectrum cutoff is determined by $m_c$. 
In our simulations, this happens at $t\approx 26\tau_A$, with no appreciable dependence on 
collisionality ($M$), because the decrease of $\gamma$ is fast [see \fig{fig:rec_rate}(a)].  
The inset in \fig{fig:mpeak_scaling} shows the time lag between the peaks of the reconnection 
and dissipation rates $vs.$ $M$. The logarithmic dependence 
is due to the fast decay of $\gamma$, and the consequent rapid increase of 
$m_\gamma\sim\gamma^{-2}$ to overtake $m_c$, thus enabling dissipation.
The weak dependence of the lag on collisions 
implies that dissipation occurs in 
finite time even for weak collisionality.

The value of $m=m_{peak}$ at which most energy is dissipated is the solution of
$d (\hypercoll m^4 E_m)/dm=0$, with $E_m$ given by \eq{eq:E_m} in the regime $m_\gamma\gg m_c$. 
This yields $m_{peak}=(9/7)^{2/9}m_c$. 
This expression, evaluated for 
$k_y=1$~\footnote{We chose $k_y=1$ because 
this is the characteristic scale of the energy transfer, as is clear from the 
top row of \fig{fig:spectrum}. Note, however, that $m_c\sim k_y^{2/9}$, a fairly weak dependence.},
is compared  in \fig{fig:mpeak_scaling} with the numerically 
determined value of $m$ at which the dissipation peaks. 
The remarkable agreement that is obtained shows
that the electron heating we observe is the result of {\it linear} phase mixing.

The phase mixing process discussed above is illustrated by the plots in 
the bottom row of~\fig{fig:spectrum} showing the electron distribution function $g_e$, 
taken at $y/a=2.3$ 
(i.e., inside the island; see \fig{contour}). The progressive creation of 
finer scales in velocity-space, a textbook signature of phase mixing, 
is manifest (cf.~\cite{Grasso_01,Liseikina_04}).
\paragraph{Conclusions.}
This Letter presents the first detailed investigations of electron heating caused
by magnetic reconnection in strongly magnetized, weakly collisional plasmas.
Using a novel fluid-kinetic framework~\cite{ZS_11}, 
we were able to show that
linear phase mixing/Landau damping 
is the main mechanism for energy conversion and electron heating. 
Reconnection and electron heating are causally related, but temporally 
and spatially disconnected: heating happens after most flux has reconnected, and along the 
island separatrices, not in the current sheet.
Our other key conclusions are: 
(i) the maximum reconnection rate in the asymptotically-large-guide-field 
limit is as fast as in the no-guide-field limit, $cE_{max}\sim 0.2 v_AB_{y,0}$, provided that 
the system is large enough; (ii) the saturation amplitude
in the kinetic (weakly collisional)
regime is identical to that in MHD (collisional) systems~\cite{Loureiro_05}, 
as long as the island is large compared to the kinetic scales.
The electron inertia scale appears to provide the lower boundary on the 
saturation amplitude --- this result may be important to the understanding 
of magnetized turbulence (e.g.,~\cite{Doerk_11,Guttenfelder_11,Nevins_11,Hatch_12}), 
as it effectively sets the minimum fluctuation amplitude.

\paragraph{Acknowledgments.}
This work was supported by 
Funda\c{c}\~ao para a Ci\^{e}ncia e Tecnologia (Ci\^encia 2008 and 
Grant no. PTDC/FIS/118187/2010), by the European Communities
under the contracts of Association between EURATOM and IST
and EURATOM and CCFE, and by the Leverhulme Trust 
Network for Magnetised Plasma Turbulence.
The views and opinions expressed herein do not necessarily reflect those of the 
European Commission.
Simulations were carried out at HPC-FF (Juelich), Jugene (PRACE) and
Ranger (NCSA).



\begin{thebibliography}{39}%
\makeatletter
\providecommand \@ifxundefined [1]{%
 \@ifx{#1\undefined}
}%
\providecommand \@ifnum [1]{%
 \ifnum #1\expandafter \@firstoftwo
 \else \expandafter \@secondoftwo
 \fi
}%
\providecommand \@ifx [1]{%
 \ifx #1\expandafter \@firstoftwo
 \else \expandafter \@secondoftwo
 \fi
}%
\providecommand \natexlab [1]{#1}%
\providecommand \enquote  [1]{``#1''}%
\providecommand \bibnamefont  [1]{#1}%
\providecommand \bibfnamefont [1]{#1}%
\providecommand \citenamefont [1]{#1}%
\providecommand \href@noop [0]{\@secondoftwo}%
\providecommand \href [0]{\begingroup \@sanitize@url \@href}%
\providecommand \@href[1]{\@@startlink{#1}\@@href}%
\providecommand \@@href[1]{\endgroup#1\@@endlink}%
\providecommand \@sanitize@url [0]{\catcode `\\12\catcode `\$12\catcode
  `\&12\catcode `\#12\catcode `\^12\catcode `\_12\catcode `\%12\relax}%
\providecommand \@@startlink[1]{}%
\providecommand \@@endlink[0]{}%
\providecommand \url  [0]{\begingroup\@sanitize@url \@url }%
\providecommand \@url [1]{\endgroup\@href {#1}{\urlprefix }}%
\providecommand \urlprefix  [0]{URL }%
\providecommand \Eprint [0]{\href }%
\providecommand \doibase [0]{http://dx.doi.org/}%
\providecommand \selectlanguage [0]{\@gobble}%
\providecommand \bibinfo  [0]{\@secondoftwo}%
\providecommand \bibfield  [0]{\@secondoftwo}%
\providecommand \translation [1]{[#1]}%
\providecommand \BibitemOpen [0]{}%
\providecommand \bibitemStop [0]{}%
\providecommand \bibitemNoStop [0]{.\EOS\space}%
\providecommand \EOS [0]{\spacefactor3000\relax}%
\providecommand \BibitemShut  [1]{\csname bibitem#1\endcsname}%
\let\auto@bib@innerbib\@empty
\bibitem [{\citenamefont {Zweibel}\ and\ \citenamefont
  {Yamada}(2009)}]{Zweibel_09}%
  \BibitemOpen
  \bibfield  {author} {\bibinfo {author} {\bibfnamefont {E.~G.}\ \bibnamefont
  {Zweibel}}\ and\ \bibinfo {author} {\bibfnamefont {M.}~\bibnamefont
  {Yamada}},\ }\href@noop {} {\bibfield  {journal} {\bibinfo  {journal} {Ann.
  Rev. Astron. Astrophys.}\ }\textbf {\bibinfo {volume} {47}},\ \bibinfo
  {pages} {291} (\bibinfo {year} {2009})}\BibitemShut {NoStop}%
\bibitem [{\citenamefont {{Furth}}\ \emph {et~al.}(1963)\citenamefont
  {{Furth}}, \citenamefont {{Killeen}},\ and\ \citenamefont
  {{Rosenbluth}}}]{FKR}%
  \BibitemOpen
  \bibfield  {author} {\bibinfo {author} {\bibfnamefont {H.~P.}\ \bibnamefont
  {{Furth}}}, \bibinfo {author} {\bibfnamefont {J.}~\bibnamefont {{Killeen}}},
  \ and\ \bibinfo {author} {\bibfnamefont {M.~N.}\ \bibnamefont
  {{Rosenbluth}}},\ }\href {\doibase 10.1063/1.1706761} {\bibfield  {journal}
  {\bibinfo  {journal} {Phys. Fluids}\ }\textbf {\bibinfo {volume} {6}},\
  \bibinfo {pages} {459} (\bibinfo {year} {1963})}\BibitemShut {NoStop}%
\bibitem [{\citenamefont {{Rutherford}}(1973)}]{Rutherford_73}%
  \BibitemOpen
  \bibfield  {author} {\bibinfo {author} {\bibfnamefont {P.~H.}\ \bibnamefont
  {{Rutherford}}},\ }\href {\doibase 10.1063/1.1694232} {\bibfield  {journal}
  {\bibinfo  {journal} {Phys. Fluids}\ }\textbf {\bibinfo {volume} {16}},\
  \bibinfo {pages} {1903} (\bibinfo {year} {1973})}\BibitemShut {NoStop}%
\bibitem [{\citenamefont {{Escande}}\ and\ \citenamefont
  {{Ottaviani}}(2004)}]{Escande_04}%
  \BibitemOpen
  \bibfield  {author} {\bibinfo {author} {\bibfnamefont {D.~F.}\ \bibnamefont
  {{Escande}}}\ and\ \bibinfo {author} {\bibfnamefont {M.}~\bibnamefont
  {{Ottaviani}}},\ }\href@noop {} {\bibfield  {journal} {\bibinfo  {journal}
  {Phys. Lett. A}\ }\textbf {\bibinfo {volume} {323}},\ \bibinfo {pages} {278}
  (\bibinfo {year} {2004})}\BibitemShut {NoStop}%
\bibitem [{\citenamefont {{Militello}}\ and\ \citenamefont
  {{Porcelli}}(2004)}]{Militello_04}%
  \BibitemOpen
  \bibfield  {author} {\bibinfo {author} {\bibfnamefont {F.}~\bibnamefont
  {{Militello}}}\ and\ \bibinfo {author} {\bibfnamefont {F.}~\bibnamefont
  {{Porcelli}}},\ }\href@noop {} {\bibfield  {journal} {\bibinfo  {journal}
  {Phys. Plasmas}\ }\textbf {\bibinfo {volume} {11}},\ \bibinfo {pages} {L13}
  (\bibinfo {year} {2004})}\BibitemShut {NoStop}%
\bibitem [{\citenamefont {{Loureiro}}\ \emph {et~al.}(2005)\citenamefont
  {{Loureiro}}, \citenamefont {{Cowley}}, \citenamefont {{Dorland}},
  \citenamefont {{Haines}},\ and\ \citenamefont
  {{Schekochihin}}}]{Loureiro_05}%
  \BibitemOpen
  \bibfield  {author} {\bibinfo {author} {\bibfnamefont {N.~F.}\ \bibnamefont
  {{Loureiro}}}, \bibinfo {author} {\bibfnamefont {S.~C.}\ \bibnamefont
  {{Cowley}}}, \bibinfo {author} {\bibfnamefont {W.~D.}\ \bibnamefont
  {{Dorland}}}, \bibinfo {author} {\bibfnamefont {M.~G.}\ \bibnamefont
  {{Haines}}}, \ and\ \bibinfo {author} {\bibfnamefont {A.~A.}\ \bibnamefont
  {{Schekochihin}}},\ }\href {\doibase 10.1103/PhysRevLett.95.235003}
  {\bibfield  {journal} {\bibinfo  {journal} {Phys. Rev. Lett.}\ }\textbf
  {\bibinfo {volume} {95}},\ \bibinfo {eid} {235003} (\bibinfo {year}
  {2005})}\BibitemShut {NoStop}%
\bibitem [{\citenamefont {{Loureiro}}\ \emph {et~al.}(2012)\citenamefont
  {{Loureiro}}, \citenamefont {{Samtaney}}, \citenamefont {{Schekochihin}},\
  and\ \citenamefont {{Uzdensky}}}]{Loureiro_12}%
  \BibitemOpen
  \bibfield  {author} {\bibinfo {author} {\bibfnamefont {N.~F.}\ \bibnamefont
  {{Loureiro}}}, \bibinfo {author} {\bibfnamefont {R.}~\bibnamefont
  {{Samtaney}}}, \bibinfo {author} {\bibfnamefont {A.~A.}\ \bibnamefont
  {{Schekochihin}}}, \ and\ \bibinfo {author} {\bibfnamefont {D.~A.}\
  \bibnamefont {{Uzdensky}}},\ }\href {\doibase 10.1063/1.3703318} {\bibfield
  {journal} {\bibinfo  {journal} {Phys. Plasmas}\ }\textbf {\bibinfo {volume}
  {19}},\ \bibinfo {pages} {042303} (\bibinfo {year} {2012})}\BibitemShut
  {NoStop}%
\bibitem [{\citenamefont {{Zocco}}\ and\ \citenamefont
  {{Schekochihin}}(2011)}]{ZS_11}%
  \BibitemOpen
  \bibfield  {author} {\bibinfo {author} {\bibfnamefont {A.}~\bibnamefont
  {{Zocco}}}\ and\ \bibinfo {author} {\bibfnamefont {A.~A.}\ \bibnamefont
  {{Schekochihin}}},\ }\href {\doibase 10.1063/1.3628639} {\bibfield  {journal}
  {\bibinfo  {journal} {Phys. Plasmas}\ }\textbf {\bibinfo {volume} {18}},\
  \bibinfo {pages} {102309} (\bibinfo {year} {2011})}\BibitemShut {NoStop}%
\bibitem [{\citenamefont {{Frieman}}\ and\ \citenamefont
  {{Chen}}(1982)}]{Frieman_Chen_82}%
  \BibitemOpen
  \bibfield  {author} {\bibinfo {author} {\bibfnamefont {E.~A.}\ \bibnamefont
  {{Frieman}}}\ and\ \bibinfo {author} {\bibfnamefont {L.}~\bibnamefont
  {{Chen}}},\ }\href {\doibase 10.1063/1.863762} {\bibfield  {journal}
  {\bibinfo  {journal} {Phys. Fluids}\ }\textbf {\bibinfo {volume} {25}},\
  \bibinfo {pages} {502} (\bibinfo {year} {1982})}\BibitemShut {NoStop}%
\bibitem [{\citenamefont {{Krommes}}(2002)}]{Krommes_02}%
  \BibitemOpen
  \bibfield  {author} {\bibinfo {author} {\bibfnamefont {J.~A.}\ \bibnamefont
  {{Krommes}}},\ }\href {\doibase 10.1016/S0370-1573(01)00066-7} {\bibfield
  {journal} {\bibinfo  {journal} {Phys. Rep.}\ }\textbf {\bibinfo {volume}
  {360}},\ \bibinfo {pages} {1} (\bibinfo {year} {2002})}\BibitemShut {NoStop}%
\bibitem [{\citenamefont {{Howes}}\ \emph {et~al.}(2006)\citenamefont
  {{Howes}}, \citenamefont {{Cowley}}, \citenamefont {{Dorland}}, \citenamefont
  {{Hammett}}, \citenamefont {{Quataert}},\ and\ \citenamefont
  {{Schekochihin}}}]{Howes_06}%
  \BibitemOpen
  \bibfield  {author} {\bibinfo {author} {\bibfnamefont {G.~G.}\ \bibnamefont
  {{Howes}}}, \bibinfo {author} {\bibfnamefont {S.~C.}\ \bibnamefont
  {{Cowley}}}, \bibinfo {author} {\bibfnamefont {W.}~\bibnamefont {{Dorland}}},
  \bibinfo {author} {\bibfnamefont {G.~W.}\ \bibnamefont {{Hammett}}}, \bibinfo
  {author} {\bibfnamefont {E.}~\bibnamefont {{Quataert}}}, \ and\ \bibinfo
  {author} {\bibfnamefont {A.~A.}\ \bibnamefont {{Schekochihin}}},\ }\href
  {\doibase 10.1086/506172} {\bibfield  {journal} {\bibinfo  {journal}
  {Astrophys. J.}\ }\textbf {\bibinfo {volume} {651}},\ \bibinfo {pages} {590}
  (\bibinfo {year} {2006})}\BibitemShut {NoStop}%
\bibitem [{\citenamefont {{Abel}}\ \emph {et~al.}(2012)\citenamefont {{Abel}},
  \citenamefont {{Plunk}}, \citenamefont {{Wang}}, \citenamefont {{Barnes}},
  \citenamefont {{Cowley}}, \citenamefont {{Dorland}},\ and\ \citenamefont
  {{Schekochihin}}}]{Abel_12}%
  \BibitemOpen
  \bibfield  {author} {\bibinfo {author} {\bibfnamefont {I.~G.}\ \bibnamefont
  {{Abel}}}, \bibinfo {author} {\bibfnamefont {G.~G.}\ \bibnamefont {{Plunk}}},
  \bibinfo {author} {\bibfnamefont {E.}~\bibnamefont {{Wang}}}, \bibinfo
  {author} {\bibfnamefont {M.}~\bibnamefont {{Barnes}}}, \bibinfo {author}
  {\bibfnamefont {S.~C.}\ \bibnamefont {{Cowley}}}, \bibinfo {author}
  {\bibfnamefont {W.}~\bibnamefont {{Dorland}}}, \ and\ \bibinfo {author}
  {\bibfnamefont {A.~A.}\ \bibnamefont {{Schekochihin}}},\ }\href@noop {}
  {\bibfield  {journal} {\bibinfo  {journal} {ArXiv e-prints}\ } (\bibinfo
  {year} {2012})},\ \Eprint {http://arxiv.org/abs/1209.4782} {arXiv:1209.4782
  [physics.plasm-ph]} \BibitemShut {NoStop}%
\bibitem [{Note1()}]{Note1}%
  \BibitemOpen
  \bibinfo {note} {We keep the collision operator in Eq.~(\ref {ge_eq}), but
  ignore the resistive term $\eta \nabla _\perp ^2A_{\parallel }$ on the RHS of
  Eq.~(\ref {eq:Ohm}). This is formally inconsistent, since $\eta = \nu
  _{ei}d_e^2$; in practice, the only effect of setting $\eta =0$ is allowing
  the use of larger values of $\nu _{ei}$ than would otherwise be required in
  order to access the collisionless regime.}\BibitemShut {Stop}%
\bibitem [{\citenamefont {{Schep}}\ \emph {et~al.}(1994)\citenamefont
  {{Schep}}, \citenamefont {{Pegoraro}},\ and\ \citenamefont
  {{Kuvshinov}}}]{Schep_94}%
  \BibitemOpen
  \bibfield  {author} {\bibinfo {author} {\bibfnamefont {T.~J.}\ \bibnamefont
  {{Schep}}}, \bibinfo {author} {\bibfnamefont {F.}~\bibnamefont {{Pegoraro}}},
  \ and\ \bibinfo {author} {\bibfnamefont {B.~N.}\ \bibnamefont
  {{Kuvshinov}}},\ }\href {\doibase 10.1063/1.870523} {\bibfield  {journal}
  {\bibinfo  {journal} {Phys. Plasmas}\ }\textbf {\bibinfo {volume} {1}},\
  \bibinfo {pages} {2843} (\bibinfo {year} {1994})}\BibitemShut {NoStop}%
\bibitem [{\citenamefont {{Loureiro}}\ and\ \citenamefont
  {{Hammett}}(2008)}]{Loureiro_08}%
  \BibitemOpen
  \bibfield  {author} {\bibinfo {author} {\bibfnamefont {N.~F.}\ \bibnamefont
  {{Loureiro}}}\ and\ \bibinfo {author} {\bibfnamefont {G.~W.}\ \bibnamefont
  {{Hammett}}},\ }\href@noop {} {\bibfield  {journal} {\bibinfo  {journal} {J.
  Comp. Phys.}\ }\textbf {\bibinfo {volume} {227}},\ \bibinfo {pages} {4518}
  (\bibinfo {year} {2008})}\BibitemShut {NoStop}%
\bibitem [{\citenamefont {Ottaviani}\ and\ \citenamefont
  {Porcelli}(1993)}]{Ottaviani_93}%
  \BibitemOpen
  \bibfield  {author} {\bibinfo {author} {\bibfnamefont {M.}~\bibnamefont
  {Ottaviani}}\ and\ \bibinfo {author} {\bibfnamefont {F.}~\bibnamefont
  {Porcelli}},\ }\href {\doibase 10.1103/PhysRevLett.71.3802} {\bibfield
  {journal} {\bibinfo  {journal} {Phys. Rev. Lett.}\ }\textbf {\bibinfo
  {volume} {71}},\ \bibinfo {pages} {3802} (\bibinfo {year}
  {1993})}\BibitemShut {NoStop}%
\bibitem [{\citenamefont {{Rogers}}\ and\ \citenamefont
  {{Zakharov}}(1996)}]{Rogers_96}%
  \BibitemOpen
  \bibfield  {author} {\bibinfo {author} {\bibfnamefont {B.}~\bibnamefont
  {{Rogers}}}\ and\ \bibinfo {author} {\bibfnamefont {L.}~\bibnamefont
  {{Zakharov}}},\ }\href {\doibase 10.1063/1.871925} {\bibfield  {journal}
  {\bibinfo  {journal} {Phys. Plasmas}\ }\textbf {\bibinfo {volume} {3}},\
  \bibinfo {pages} {2411} (\bibinfo {year} {1996})}\BibitemShut {NoStop}%
\bibitem [{\citenamefont {{Valori}}\ \emph {et~al.}(2000)\citenamefont
  {{Valori}}, \citenamefont {{Grasso}},\ and\ \citenamefont {{de
  Blank}}}]{Valori_00}%
  \BibitemOpen
  \bibfield  {author} {\bibinfo {author} {\bibfnamefont {G.}~\bibnamefont
  {{Valori}}}, \bibinfo {author} {\bibfnamefont {D.}~\bibnamefont {{Grasso}}},
  \ and\ \bibinfo {author} {\bibfnamefont {H.~J.}\ \bibnamefont {{de Blank}}},\
  }\href@noop {} {\bibfield  {journal} {\bibinfo  {journal} {Phys. Plasmas}\
  }\textbf {\bibinfo {volume} {7}},\ \bibinfo {pages} {178} (\bibinfo {year}
  {2000})}\BibitemShut {NoStop}%
\bibitem [{\citenamefont {{Birn}}\ \emph {et~al.}(2001)\citenamefont {{Birn}}
  \emph {et~al.}}]{Birn_01}%
  \BibitemOpen
  \bibfield  {author} {\bibinfo {author} {\bibfnamefont {J.}~\bibnamefont
  {{Birn}}} \emph {et~al.},\ }\href@noop {} {\bibfield  {journal} {\bibinfo
  {journal} {J. Geophys. Res.}\ }\textbf {\bibinfo {volume} {106}},\ \bibinfo
  {pages} {3715} (\bibinfo {year} {2001})}\BibitemShut {NoStop}%
\bibitem [{\citenamefont {{Rogers}}\ \emph {et~al.}(2001)\citenamefont
  {{Rogers}}, \citenamefont {{Denton}}, \citenamefont {{Drake}},\ and\
  \citenamefont {{Shay}}}]{Rogers_01}%
  \BibitemOpen
  \bibfield  {author} {\bibinfo {author} {\bibfnamefont {B.~N.}\ \bibnamefont
  {{Rogers}}}, \bibinfo {author} {\bibfnamefont {R.~E.}\ \bibnamefont
  {{Denton}}}, \bibinfo {author} {\bibfnamefont {J.~F.}\ \bibnamefont
  {{Drake}}}, \ and\ \bibinfo {author} {\bibfnamefont {M.~A.}\ \bibnamefont
  {{Shay}}},\ }\href {\doibase 10.1103/PhysRevLett.87.195004} {\bibfield
  {journal} {\bibinfo  {journal} {Phys. Rev. Lett.}\ }\textbf {\bibinfo
  {volume} {87}},\ \bibinfo {eid} {195004} (\bibinfo {year}
  {2001})}\BibitemShut {NoStop}%
\bibitem [{\citenamefont {{Rogers}}\ \emph {et~al.}(2007)\citenamefont
  {{Rogers}}, \citenamefont {{Kobayashi}}, \citenamefont {{Ricci}},
  \citenamefont {{Dorland}}, \citenamefont {{Drake}},\ and\ \citenamefont
  {{Tatsuno}}}]{rogers_07}%
  \BibitemOpen
  \bibfield  {author} {\bibinfo {author} {\bibfnamefont {B.~N.}\ \bibnamefont
  {{Rogers}}}, \bibinfo {author} {\bibfnamefont {S.}~\bibnamefont
  {{Kobayashi}}}, \bibinfo {author} {\bibfnamefont {P.}~\bibnamefont
  {{Ricci}}}, \bibinfo {author} {\bibfnamefont {W.}~\bibnamefont {{Dorland}}},
  \bibinfo {author} {\bibfnamefont {J.}~\bibnamefont {{Drake}}}, \ and\
  \bibinfo {author} {\bibfnamefont {T.}~\bibnamefont {{Tatsuno}}},\ }\href
  {\doibase 10.1063/1.2774003} {\bibfield  {journal} {\bibinfo  {journal}
  {Phys. Plasmas}\ }\textbf {\bibinfo {volume} {14}},\ \bibinfo {pages}
  {092110} (\bibinfo {year} {2007})}\BibitemShut {NoStop}%
\bibitem [{\citenamefont {{Uzdensky}}\ and\ \citenamefont
  {{Kulsrud}}(2006)}]{Uzdensky_06}%
  \BibitemOpen
  \bibfield  {author} {\bibinfo {author} {\bibfnamefont {D.~A.}\ \bibnamefont
  {{Uzdensky}}}\ and\ \bibinfo {author} {\bibfnamefont {R.~M.}\ \bibnamefont
  {{Kulsrud}}},\ }\href {\doibase 10.1063/1.2209627} {\bibfield  {journal}
  {\bibinfo  {journal} {Phys. Plasmas}\ }\textbf {\bibinfo {volume} {13}},\
  \bibinfo {pages} {062305} (\bibinfo {year} {2006})}\BibitemShut {NoStop}%
\bibitem [{Note2()}]{Note2}%
  \BibitemOpen
  \bibinfo {note} {Note that in none of our runs did we observe the current
  sheet becoming unstable to plasmoid formation, though this may be because
  even our largest system is in fact not large enough: the longest box is
  $L_y/a=3.785\pi $ (for $\Delta 'a=54$); this run yields the maximum ratio of
  the current sheet length to $\rho _s$ of $\sim 8$, too small to expect a
  transition to multiple X-line reconnection~\cite {Ji_Daughton_11,
  Huang_11}.}\BibitemShut {Stop}%
\bibitem [{\citenamefont {{Sydora}}(2001)}]{Sydora_01}%
  \BibitemOpen
  \bibfield  {author} {\bibinfo {author} {\bibfnamefont {R.~D.}\ \bibnamefont
  {{Sydora}}},\ }\href {\doibase 10.1063/1.1362536} {\bibfield  {journal}
  {\bibinfo  {journal} {Phys. Plasmas}\ }\textbf {\bibinfo {volume} {8}},\
  \bibinfo {pages} {1929} (\bibinfo {year} {2001})}\BibitemShut {NoStop}%
\bibitem [{Note3()}]{Note3}%
  \BibitemOpen
  \bibinfo {note} {If we use a (formally incorrect) isothermal closure, most of
  the dissipation occurs via Ohmic heating.}\BibitemShut {Stop}%
\bibitem [{Note4()}]{Note4}%
  \BibitemOpen
  \bibinfo {note} {This is why the reconnection rate should be independent of
  the details of the dissipation: indeed, the same rate should be obtained even
  in dissipation-free Hamiltonian models~\cite
  {Kuvshinov_94,Grasso_01,Tassi_08,Waelbroeck_09,Sarto_11}, although they would
  not obtain the same energy partition that we report here.}\BibitemShut
  {Stop}%
\bibitem [{Note5()}]{Note5}%
  \BibitemOpen
  \bibinfo {note} {We chose $k_y=1$ because this is the characteristic scale of
  the energy transfer, as is clear from the top row of Fig.~\ref
  {fig:spectrum}. Note, however, that $m_c\sim k_y^{2/9}$, a fairly weak
  dependence.}\BibitemShut {Stop}%
\bibitem [{\citenamefont {{Grasso}}\ \emph {et~al.}(2001)\citenamefont
  {{Grasso}}, \citenamefont {{Califano}}, \citenamefont {{Pegoraro}},\ and\
  \citenamefont {{Porcelli}}}]{Grasso_01}%
  \BibitemOpen
  \bibfield  {author} {\bibinfo {author} {\bibfnamefont {D.}~\bibnamefont
  {{Grasso}}}, \bibinfo {author} {\bibfnamefont {F.}~\bibnamefont
  {{Califano}}}, \bibinfo {author} {\bibfnamefont {F.}~\bibnamefont
  {{Pegoraro}}}, \ and\ \bibinfo {author} {\bibfnamefont {F.}~\bibnamefont
  {{Porcelli}}},\ }\href {\doibase 10.1103/PhysRevLett.86.5051} {\bibfield
  {journal} {\bibinfo  {journal} {Phys. Rev. Lett.}\ }\textbf {\bibinfo
  {volume} {86}},\ \bibinfo {pages} {5051} (\bibinfo {year}
  {2001})}\BibitemShut {NoStop}%
\bibitem [{\citenamefont {{Liseikina}}\ \emph {et~al.}(2004)\citenamefont
  {{Liseikina}}, \citenamefont {{Pegoraro}},\ and\ \citenamefont
  {{Echkina}}}]{Liseikina_04}%
  \BibitemOpen
  \bibfield  {author} {\bibinfo {author} {\bibfnamefont {T.~V.}\ \bibnamefont
  {{Liseikina}}}, \bibinfo {author} {\bibfnamefont {F.}~\bibnamefont
  {{Pegoraro}}}, \ and\ \bibinfo {author} {\bibfnamefont {E.~Y.}\ \bibnamefont
  {{Echkina}}},\ }\href {\doibase 10.1063/1.1758231} {\bibfield  {journal}
  {\bibinfo  {journal} {Phys. Plasmas}\ }\textbf {\bibinfo {volume} {11}},\
  \bibinfo {pages} {3535} (\bibinfo {year} {2004})}\BibitemShut {NoStop}%
\bibitem [{\citenamefont {{Doerk}}\ \emph {et~al.}(2011)\citenamefont
  {{Doerk}}, \citenamefont {{Jenko}}, \citenamefont {{Pueschel}},\ and\
  \citenamefont {{Hatch}}}]{Doerk_11}%
  \BibitemOpen
  \bibfield  {author} {\bibinfo {author} {\bibfnamefont {H.}~\bibnamefont
  {{Doerk}}}, \bibinfo {author} {\bibfnamefont {F.}~\bibnamefont {{Jenko}}},
  \bibinfo {author} {\bibfnamefont {M.~J.}\ \bibnamefont {{Pueschel}}}, \ and\
  \bibinfo {author} {\bibfnamefont {D.~R.}\ \bibnamefont {{Hatch}}},\ }\href
  {\doibase 10.1103/PhysRevLett.106.155003} {\bibfield  {journal} {\bibinfo
  {journal} {Phys. Rev. Lett.}\ }\textbf {\bibinfo {volume} {106}},\ \bibinfo
  {eid} {155003} (\bibinfo {year} {2011})}\BibitemShut {NoStop}%
\bibitem [{\citenamefont {Guttenfelder}\ \emph {et~al.}(2011)\citenamefont
  {Guttenfelder} \emph {et~al.}}]{Guttenfelder_11}%
  \BibitemOpen
  \bibfield  {author} {\bibinfo {author} {\bibfnamefont {W.}~\bibnamefont
  {Guttenfelder}} \emph {et~al.},\ }\href@noop {} {\bibfield  {journal}
  {\bibinfo  {journal} {Phys. Rev. Lett.}\ }\textbf {\bibinfo {volume} {106}},\
  \bibinfo {pages} {155004} (\bibinfo {year} {2011})}\BibitemShut {NoStop}%
\bibitem [{\citenamefont {Nevins}\ \emph {et~al.}(2011)\citenamefont {Nevins},
  \citenamefont {Wang},\ and\ \citenamefont {Candy}}]{Nevins_11}%
  \BibitemOpen
  \bibfield  {author} {\bibinfo {author} {\bibfnamefont {W.~M.}\ \bibnamefont
  {Nevins}}, \bibinfo {author} {\bibfnamefont {E.}~\bibnamefont {Wang}}, \ and\
  \bibinfo {author} {\bibfnamefont {J.}~\bibnamefont {Candy}},\ }\href
  {\doibase 10.1103/PhysRevLett.106.065003} {\bibfield  {journal} {\bibinfo
  {journal} {Phys. Rev. Lett.}\ }\textbf {\bibinfo {volume} {106}},\ \bibinfo
  {pages} {065003} (\bibinfo {year} {2011})}\BibitemShut {NoStop}%
\bibitem [{\citenamefont {Hatch}\ \emph {et~al.}(2012)\citenamefont {Hatch},
  \citenamefont {Pueschel}, \citenamefont {Jenko}, \citenamefont {Nevins},
  \citenamefont {Terry},\ and\ \citenamefont {Doerk}}]{Hatch_12}%
  \BibitemOpen
  \bibfield  {author} {\bibinfo {author} {\bibfnamefont {D.~R.}\ \bibnamefont
  {Hatch}}, \bibinfo {author} {\bibfnamefont {M.~J.}\ \bibnamefont {Pueschel}},
  \bibinfo {author} {\bibfnamefont {F.}~\bibnamefont {Jenko}}, \bibinfo
  {author} {\bibfnamefont {W.~M.}\ \bibnamefont {Nevins}}, \bibinfo {author}
  {\bibfnamefont {P.~W.}\ \bibnamefont {Terry}}, \ and\ \bibinfo {author}
  {\bibfnamefont {H.}~\bibnamefont {Doerk}},\ }\href {\doibase
  10.1103/PhysRevLett.108.235002} {\bibfield  {journal} {\bibinfo  {journal}
  {Phys. Rev. Lett.}\ }\textbf {\bibinfo {volume} {108}},\ \bibinfo {pages}
  {235002} (\bibinfo {year} {2012})}\BibitemShut {NoStop}%
\bibitem [{\citenamefont {{Ji}}\ and\ \citenamefont
  {{Daughton}}(2011)}]{Ji_Daughton_11}%
  \BibitemOpen
  \bibfield  {author} {\bibinfo {author} {\bibfnamefont {H.}~\bibnamefont
  {{Ji}}}\ and\ \bibinfo {author} {\bibfnamefont {W.}~\bibnamefont
  {{Daughton}}},\ }\href {\doibase 10.1063/1.3647505} {\bibfield  {journal}
  {\bibinfo  {journal} {Phys. Plasmas}\ }\textbf {\bibinfo {volume} {18}},\
  \bibinfo {pages} {111207} (\bibinfo {year} {2011})}\BibitemShut {NoStop}%
\bibitem [{\citenamefont {{Huang}}\ \emph {et~al.}(2011)\citenamefont
  {{Huang}}, \citenamefont {{Bhattacharjee}},\ and\ \citenamefont
  {{Sullivan}}}]{Huang_11}%
  \BibitemOpen
  \bibfield  {author} {\bibinfo {author} {\bibfnamefont {Y.-M.}\ \bibnamefont
  {{Huang}}}, \bibinfo {author} {\bibfnamefont {A.}~\bibnamefont
  {{Bhattacharjee}}}, \ and\ \bibinfo {author} {\bibfnamefont {B.~P.}\
  \bibnamefont {{Sullivan}}},\ }\href {\doibase 10.1063/1.3606363} {\bibfield
  {journal} {\bibinfo  {journal} {Phys. Plasmas}\ }\textbf {\bibinfo {volume}
  {18}},\ \bibinfo {pages} {072109} (\bibinfo {year} {2011})}\BibitemShut
  {NoStop}%
\bibitem [{\citenamefont {{Kuvshinov}}\ \emph {et~al.}(1994)\citenamefont
  {{Kuvshinov}}, \citenamefont {{Pegoraro}},\ and\ \citenamefont
  {{Schep}}}]{Kuvshinov_94}%
  \BibitemOpen
  \bibfield  {author} {\bibinfo {author} {\bibfnamefont {B.~N.}\ \bibnamefont
  {{Kuvshinov}}}, \bibinfo {author} {\bibfnamefont {F.}~\bibnamefont
  {{Pegoraro}}}, \ and\ \bibinfo {author} {\bibfnamefont {T.~J.}\ \bibnamefont
  {{Schep}}},\ }\href {\doibase 10.1016/0375-9601(94)90143-0} {\bibfield
  {journal} {\bibinfo  {journal} {Phys. Lett. A}\ }\textbf {\bibinfo {volume}
  {191}},\ \bibinfo {pages} {296} (\bibinfo {year} {1994})}\BibitemShut
  {NoStop}%
\bibitem [{\citenamefont {{Tassi}}\ \emph {et~al.}(2008)\citenamefont
  {{Tassi}}, \citenamefont {{Morrison}}, \citenamefont {{Waelbroeck}},\ and\
  \citenamefont {{Grasso}}}]{Tassi_08}%
  \BibitemOpen
  \bibfield  {author} {\bibinfo {author} {\bibfnamefont {E.}~\bibnamefont
  {{Tassi}}}, \bibinfo {author} {\bibfnamefont {P.~J.}\ \bibnamefont
  {{Morrison}}}, \bibinfo {author} {\bibfnamefont {F.~L.}\ \bibnamefont
  {{Waelbroeck}}}, \ and\ \bibinfo {author} {\bibfnamefont {D.}~\bibnamefont
  {{Grasso}}},\ }\href {\doibase 10.1088/0741-3335/50/8/085014} {\bibfield
  {journal} {\bibinfo  {journal} {Plasma Phys. Control. Fusion}\ }\textbf
  {\bibinfo {volume} {50}},\ \bibinfo {pages} {085014} (\bibinfo {year}
  {2008})}\BibitemShut {NoStop}%
\bibitem [{\citenamefont {{Waelbroeck}}\ \emph {et~al.}(2009)\citenamefont
  {{Waelbroeck}}, \citenamefont {{Hazeltine}},\ and\ \citenamefont
  {{Morrison}}}]{Waelbroeck_09}%
  \BibitemOpen
  \bibfield  {author} {\bibinfo {author} {\bibfnamefont {F.~L.}\ \bibnamefont
  {{Waelbroeck}}}, \bibinfo {author} {\bibfnamefont {R.~D.}\ \bibnamefont
  {{Hazeltine}}}, \ and\ \bibinfo {author} {\bibfnamefont {P.~J.}\ \bibnamefont
  {{Morrison}}},\ }\href {\doibase 10.1063/1.3087972} {\bibfield  {journal}
  {\bibinfo  {journal} {Phys. Plasmas}\ }\textbf {\bibinfo {volume} {16}},\
  \bibinfo {pages} {032109} (\bibinfo {year} {2009})}\BibitemShut {NoStop}%
\bibitem [{\citenamefont {{Del Sarto}}\ \emph {et~al.}(2011)\citenamefont {{Del
  Sarto}}, \citenamefont {{Marchetto}}, \citenamefont {{Pegoraro}},\ and\
  \citenamefont {{Califano}}}]{Sarto_11}%
  \BibitemOpen
  \bibfield  {author} {\bibinfo {author} {\bibfnamefont {D.}~\bibnamefont {{Del
  Sarto}}}, \bibinfo {author} {\bibfnamefont {C.}~\bibnamefont {{Marchetto}}},
  \bibinfo {author} {\bibfnamefont {F.}~\bibnamefont {{Pegoraro}}}, \ and\
  \bibinfo {author} {\bibfnamefont {F.}~\bibnamefont {{Califano}}},\ }\href
  {\doibase 10.1088/0741-3335/53/3/035008} {\bibfield  {journal} {\bibinfo
  {journal} {Plasma Phys. Control. Fusion}\ }\textbf {\bibinfo {volume} {53}},\
  \bibinfo {pages} {035008} (\bibinfo {year} {2011})}\BibitemShut {NoStop}%
\end{thebibliography}
%

\end{document}